\documentclass[a4paper,aps,preprintnumbers,showpacs,superscriptaddress,nofootinbib,amsmath,amssymb,twocolumn]{revtex4}
\usepackage{graphicx}
\usepackage{hyperref}
\usepackage{subfigure}
\usepackage{multirow}
\usepackage[toc,page]{appendix}
\usepackage[utf8]{inputenc}
\usepackage[T1]{fontenc}
\usepackage{cmap}

\def\imo{i}

\def\K{{\cal K}}

\begin{document}
\title{Quasinormal modes in higher-derivative gravity: Testing the black hole parametrization and sensitivity of overtones}
\author{R. A. Konoplya}\email{roman.konoplya@gmail.com}
\affiliation{Research Centre for Theoretical Physics and Astrophysics, \\ Institute of Physics, Silesian University in Opava, \\ Bezručovo náměstí 13, CZ-74601 Opava, Czech Republic}

\begin{abstract}
The fundamental quasinormal modes of black holes in higher-derivative gravity given by the Einstein-Weyl action are known to be moderately corrected by the Weyl term.
Here we will show that the first several overtones are highly sensitive to even a relatively small Weyl correction, which might be important when representing the earlier stage of the black hole ringdown. In addition, we have solved the problem related to the analytical parametrized approximation of the numerical black hole solution in the Einstein-Weyl theory: In some range of parameters the approximation for the metric developed up to the third order leads to the unusual highly nonmonotonic behavior of the frequencies. We have shown that this problem can be solved via the extension of the parametrization of the metric to higher orders until reaching the regime when the frequencies do not change with further increasing of the order.
\end{abstract}
\maketitle

\section{Introduction}

Proper oscillation frequencies of black holes called {\it quasinormal modes}  \cite{Kokkotas:1999bd,Berti:2009kk,Konoplya:2011qq} have been the focus of theoretical and experimental studies during the past few decades. The observation of gravitational waves from the  merger of two black holes  by the LIGO and VIRGO Collaborations \cite{Abbott:2016blz} and a series of subsequent observations \cite{LIGOScientific:2020ibl,KAGRA:2013rdx} agree with General Relativity.  Nevertheless, the angular momenta and masses of resultant black holes are determined with a large uncertainty of tens of percent. This leaves a wide window for alternative theories of gravity \cite{Konoplya:2016pmh,Yunes:2016jcc,Berti:2018vdi}. That is why quasinormal modes of the four-dimensional asymptotically flat black holes in various alternative theories of gravity have been intensively studied.

Among various alternative theories  of gravity those with higher curvature corrections are important, because they are predicted by the low-energy limit of string theory. One such theory is represented by the Einstein action corrected by the Weyl term. Then, the general vacuum four-dimensional theory up the second order in curvature is described by the following Lagrangian:
%
\begin{eqnarray} \label{lagranzian}
L &=& \sqrt{-g} (\gamma R - \alpha C_{\mu\nu \rho\sigma} C^{\mu\nu \rho\sigma} + \beta R^2),
\end{eqnarray}
where $g$ is determinant of the matrix of a metric tensor; $\alpha$, $\beta$ and $\gamma$ are constants; and $C_{\mu\nu \rho\sigma}$ is the Weyl tensor.
For spherically symmetric and asymptotically flat solutions we can choose $\gamma =1$ and $\beta =0$ without loss of generality, so that only the coupling constant $\alpha$ matters for nonrotating black holes.
The Schwarzschild solution is a solution of the Einstein-Weyl equations, yet in \cite{Lu:2015cqa}, it was shown that a non-Schwarzschild asymptotically flat solution representing a black hole is possible and a numerical solution for it was obtained. Using the general parametrization for asymptotically flat black hole spacetimes in arbitrary metric theories of gravity \cite{Rezzolla:2014mua,Konoplya:2016jvv,Younsi:2016azx}, the numerical black hole solution was approximated by an analytical expressions for the metric functions in Ref. \cite{Kokkotas:2017zwt}.
This theory and the corresponding black hole solutions were studied in a number of works \cite{Einstein-Weyl:2018pfe,QuadraticGravity:2016kip,Zinhailo:2019rwd,Zinhailo:2018ska,Konoplya:2019ppy,Wu:2019uvq}.

The asymptotically flat non-Schwarzschild solution in the Einstein-Weyl theory has an interesting property: when the Weyl coupling constant increases approaching the extremal limit (beyond which only the Schwarzschild solution survives), the black hole mass decreases, tending to zero and having the entire time, a nonvanishing radius. Therefore, the near-extremal non-Schwarzschild solution may be a ``light black hole,'' which is unlikely to be created in the course of the gravitational collapse of stars.

The low-lying quasinormal modes of non-Schwarzschild black holes in the Einstein-Weyl theory have been studied in a few works. The eikonal formula for the quasinormal modes of test fields was obtained  in Ref. \cite{Kokkotas:2017zwt}. Massless scalar field perturbations were considered in Ref. \cite{Cai:2015fia}, but as was shown in Refs. \cite{Zinhailo:2019rwd,Zinhailo:2018ska} the analysis based on the Wentzel-Kramers-Brillouin (WKB) approach and time-domain integration of the numerical solution apparently suffers from very large numerical inaccuracy, making the final results  doubtful. Indeed, Table I in Ref. \cite{Cai:2015fia} shows that the difference between the WKB and time-domain integration results reaches approximately $70 \%$. The problem consists in the insufficient precision of the numerical metric function when further taking higher derivatives and integrating the wave equation. The obtained frequencies have random, noisy-like dependence on the coupling parameter $\alpha$ (as was shown in Fig. 1 of Ref. \cite{Zinhailo:2018ska}), which turns to be far from the much more accurate values found later in Refs. \cite{Zinhailo:2019rwd,Zinhailo:2018ska}.  In Refs. \cite{Zinhailo:2019rwd,Zinhailo:2018ska} the problem was solved using the analytical approximation for the metric function \cite{Kokkotas:2017zwt}. There, the scalar, electromagnetic and Dirac low-lying quasinormal modes were studied in detail. However, despite the quasinormal frequencies this time depended smoothly upon the coupling parameter $\alpha$, one technical problem remained: at some value of $\alpha$ a highly nonmonotonic dependence, a kind of "splash," was observed for the Dirac field perturbations at lower multipole numbers $\ell$ (see Fig. 5(b) in Ref. \cite{Zinhailo:2019rwd}). In the present paper we will show that such strange behavior is not any kind of numerical inaccuracy of the calculation method for quasinormal modes, but is related to bad accuracy of the parametrization at those values of $\alpha$. The higher-order parametrization remedies this problem.

However, our main interest here is related to the first several  overtones which were not studied previously, because only the Wentzel-Kramers-Brillouin or time-domain integration methods were used for the analysis of the spectrum of the Einstein-Weyls black holes \cite{Cai:2015fia,Zinhailo:2019rwd,Zinhailo:2018ska}.
It may seem that the fundamental mode provides sufficient information about the signal; however, recently it has been shown that while the lowest mode may not feel the deformations near the event horizon of the black holes, the first few overtones may be highly sensitive to the tiny changes of the near-horizon geometry \cite{Konoplya:2022pbc,Konoplya:2022hll}.
An important recent finding was made in Ref. \cite{Giesler:2019uxc} (and later studied in Refs. \cite{Oshita:2021iyn,Forteza:2021wfq,Oshita:2022pkc}) showing that the first several overtones must be taken into account in order to model the ringdown phase, obtained by accurate numerical relativity simulations in the beginning of the quasinormal ringing and not only at the last stage. This indicates  also that the actual quasinormal ringing starts much earlier than expected. Once the overtones are taken into account, the modeling and linear profile of the ringdown are in the full concordance and allow us to extract the angular mass and momentum of the black hole. Even though it seems that the current LIGO/VIRGO observational data do not allow reliable detection of the overtones in the gravitational-wave signals \cite{Capano:2021etf,Cotesta:2022pci}, there are indications that in the future LISA experiments the overtones could be significantly excited and cannot be neglected at the early ringdown phase \cite{Oshita:2022yry}.

The paper is organized as follows: Section II is devoted to the basic equation for the metric and the wave euqation. In Sec.~III we briefly describe the boundary condition for quasinormal modes, explain the method used for their calculations and review the obtained results. In the last section we discuss a couple of open questions.

\section{Black hole metric and wave equation}\label{sec:metricsection}

\begin{figure}
\resizebox{\linewidth}{!}{\includegraphics{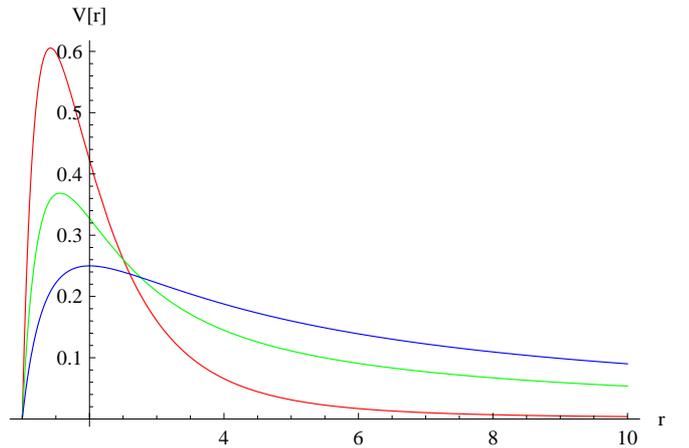}}
\caption{The effective potential for scalar field perturbations $\ell=0$: $p=1.14$ (top, red), $p=1$ (green), $p=0.876$ (bottom, blue).}
  \label{fig:potential}
\end{figure}

The general spherically symmetric metric has the form
\begin{equation}\label{metric}
ds^2 =
-A(r, p)dt^2+\dfrac{dr^2} {B(r, p)}+r^2 (\sin^2 \theta d\phi^2+d\theta^2),
\end{equation}
where the dimensionless parameter $p$ is
\begin{equation}
p=\dfrac{r_0}{\sqrt{2 \alpha}},
\end{equation}
and $r_0$ is the black hole radius.

The metric functions $A(r, p)$ and $B(r, p)$ were obtained numerically in Refs. \cite{Lu:2015cqa,Lu:2015psa}.
The analytical approximation for the black hole metric functions $A(r, p)$ and $B(r, p)$ was found in Ref. \cite{Kokkotas:2017zwt} and, as these metric functions have rather lengthy form we do not write them here explicitly and refer the reader to Ref. \cite{Kokkotas:2017zwt} instead.

The Schwarzschild solution is the exact solution of the Einstein-Weyl equations for all $p$, but the second solution also describes an asymptotically flat black hole existing at some interval $(p_{min}, p_{max})$ \cite{Lu:2015cqa,Lu:2015psa}:

\begin{subequations}
\begin{eqnarray}
p_{min}\approx\dfrac{1054}{1203}\approx0.876,\label{p_min}\\
p_{max}\approx1.14.
\end{eqnarray}
\end{subequations}

The general covariant equation for a minimally coupled scalar field has the form:
\begin{equation}\label{gensc}
\dfrac{1}{\sqrt{-g}} \partial_\mu(\sqrt{-g} g^{\mu\nu} \partial_\nu\Phi) = 0,
\end{equation}
while for an electromagnetic field the general covariant Maxwell equations read as follows:
\begin{equation}\label{genelm}
\dfrac{1}{\sqrt{-g}} \partial_\mu(F_{\rho\sigma}g^{\rho\nu}g^{\sigma\mu}\sqrt{-g})=0.
\end{equation}
Here, $A_{\mu}$ is a vector potential and $F_{\rho\sigma}=\partial_\rho A_\sigma - \partial_\sigma A_\rho$. The function $\Phi$ can be represented in terms of spherical harmonics and radial part in the following way+:
\begin{equation} \label{psi}
\Phi(t,r,y,\phi)=e^{\pm\imo\omega t}R_{\omega \ell}(r)Y_{\ell(\theta,\phi)}.
\end{equation}

Here, $Y_l(\theta,\phi)$ are the standard spherical harmonics, and $\ell$ is the multipole number. Then, we rewrite the radial component $R_{\omega \ell}(r)$ via the new function,
\begin{equation}
\Psi_{\omega l}(r)=r R_{\omega l}(r).
\end{equation}

Thus, taking into account the replacement (\ref{psi}), Eqs (\ref{gensc}) and (\ref{genelm}) can be represented in the following general form:
\begin{equation}  \label{wave-equation}
\dfrac{d^2 \Psi}{dr_*^2}+(\omega^2-V(r, p))\Psi=0,
\end{equation}
where  the ``tortoise coordinate'' $r_*$  is
\begin{equation}
dr_*=\dfrac{dr}{\sqrt{A(r, p) B(r, p)}}.
\end{equation}
The corresponding effective potentials for the scalar $V_{s}$ and electromagnetic $V_{e}$ perturbations can be written in the following forms \cite{Konoplya:2006rv}:
%
$$ V_{s}(r, p)=\dfrac{\ell(\ell+1)A(r, p)}{r^2}+\ $$
\begin{equation}
\dfrac{B(r, p)\dfrac{\partial A(r, p)}{\partial r}+A(r, p)\dfrac{\partial B (r, p)}{\partial r}}{2r},
\end{equation}
\begin{equation}
V_{e}(r, p)=\dfrac{\ell(\ell+1)A(r, p)}{r^2}.
\end{equation}
The effective potentials have the form of a positive definite potential barrier which decays at the event horizon and at infinity (see an example for the $\ell=0$ scalar field perturbations in Fig. \ref{fig:potential}).

\begin{figure*}
\resizebox{\linewidth}{!}{\includegraphics{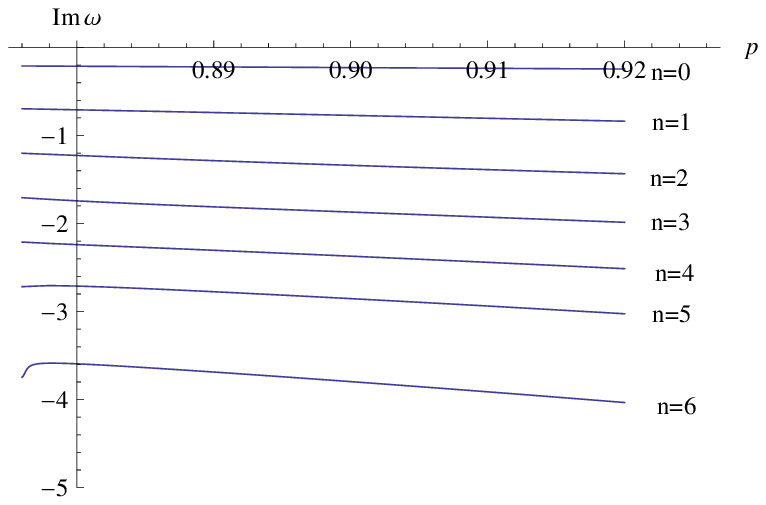}\includegraphics{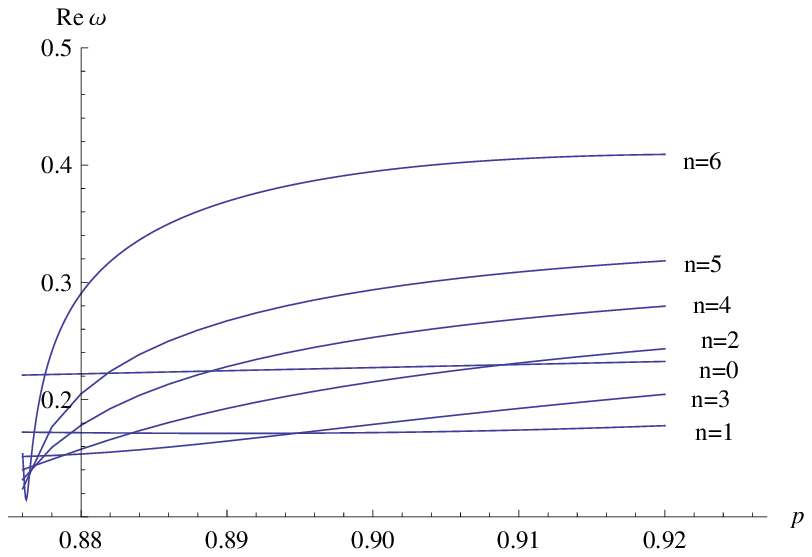}}
\caption{Real and imaginary parts of quasinormal frequencies for $\ell=0$ scalar perturbations as a function of $p$ found by the Frobenius method; $r_{+} =1$. }
  \label{fig:scal}
\end{figure*}

\begin{figure*}
\resizebox{\linewidth}{!}{\includegraphics{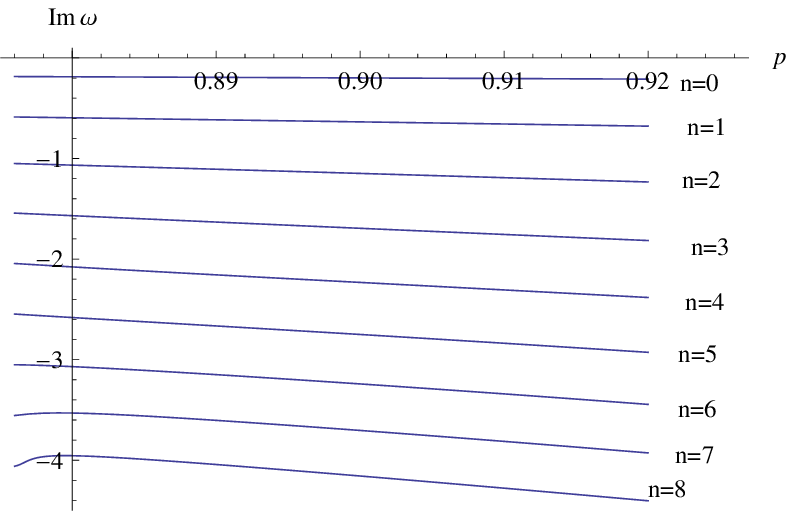}\includegraphics{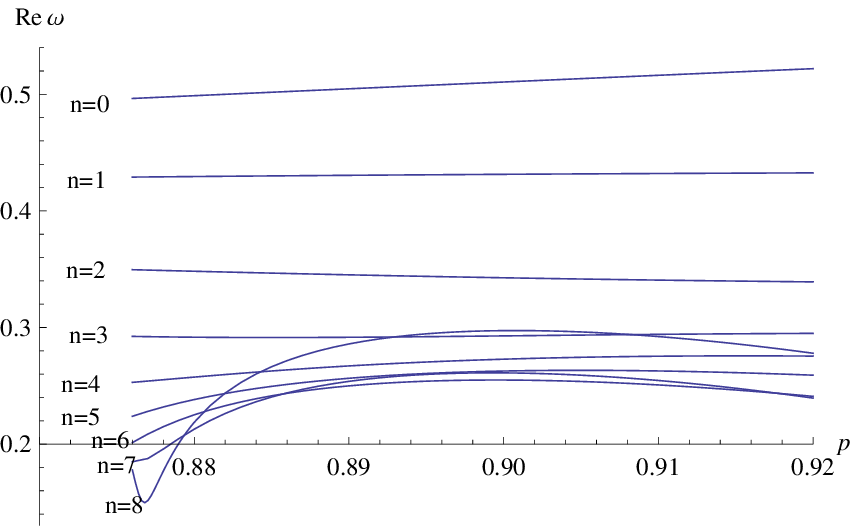}}
\caption{Real and imaginary parts of quasinormal frequencies for $\ell=1$ electromagnetic perturbations as a function of $p$ found by the Frobenius method; $r_{+} =1$}
  \label{fig:el}
\end{figure*}

\section{Quasinormal modes: sensitivity of overtones}\label{sec:scalarsection}

Quasinormal modes satisfy the following boundary conditions
\begin{equation}\label{boundary}
\Phi(r_*) \propto e^{\pm\imo\omega r_*}, \qquad r_*\to\pm\infty,
\end{equation}
corresponding to purely ingoing waves at the event horizon and purely outgoing waves at the cosmological horizon. Thus, no waves are coming from either of the boundaries, which means that a response in perturbations is detected once the source of perturbations stops acting, or, in other words, {\it proper} oscillation frequencies are observed.

In the previous literature quasinormal modes were studied with the help of either  the third-order WKB method \cite{Cai:2015fia} or the sixth-order WKB method with the Prony approximants and time-domain integration \cite{Zinhailo:2019rwd,Zinhailo:2018ska}. While the time-domain integration and the sixth-order WKB method usually  provides sufficiently good accuracy, there are some deficiencies: the WKB series converges only asymptotically and cannot\emph{ guarantee}  accuracy, while time-domain integration, especially for $\ell =0$ scalar perturbations allows us to extract values of the frequencies with only a few digits. In this respect, the Frobenius method which was applied to finding quasinormal modes by E. Leaver ~\cite{Leaver:1985ax} is based on the \emph{convergent} procedure. Therefore, it guarantees results with as many digits as one wishes. More importantly, while both WKB and time-domain integration fail when searching for overtones with $n > \ell$, the Leaver method allows one to find higher overtones accurately as well.
That is why, in order to find the accurate values of QNMs we will apply the Leaver method~\cite{Leaver:1985ax}.

\subsection{Frobenius method}

The wavelike equation~(\ref{wave-equation}) has regular singularities at the event horizon $r=r_+$ and an irregular singularity at spatial infinity $r=\infty$. Further, we will use the new function,
\begin{equation}\label{reg}
\Psi(r)=e^{\imo\omega r}r^{\lambda}\left(\frac{r-r_+}{r}\right)^{-\imo\omega/f'(r_+)}y(r),
\end{equation}
where $\lambda$ is defined in such a way that $y(r)$ is regular at $r=\infty$ once $\Psi(r)$ corresponds to the purely outgoing wave at spatial infinity. We notice that $\Psi(r)$ behaves as a purely ingoing wave at the event horizon, if $y(r)$ is regular at $r=r_+$. Therefore, we represent $y(r)$ in terms of the following Frobenius series:
\begin{equation}\label{Frobenius}
y(r)=\sum_{k=0}^{\infty}a_k\left(\frac{r-r_+}{r}\right)^k.
\end{equation}
Then, we find that the coefficients $a_k$ satisfy some recurrence relation, which can be reduced to the three-term recurrence relation via Gaussian elimination (see Ref. \cite{Konoplya:2011qq} for details). Finally, using the recurrence relation coefficients, we find the equation with an infinite continued fraction with respect to $\omega$, which is satisfied once the series in Eq. (\ref{Frobenius}) converges at $r=\infty$ i.e., when $\Psi(r)$ obeys the quasinormal boundary conditions. In order to calculate the infinite continued fraction, we also use the so called Nollert improvement~\cite{Nollert:1993zz}, which was generalized in~\cite{Zhidenko:2006rs} for an arbitrary number of terms of the recurrence relation.

\subsection{WKB method}
In the frequency domain we will use the semianalytic WKB approach applied by Will and Schutz \cite{Schutz:1985zz} for finding quasinormal modes. The Will-Schutz formula was extended to higher orders in Refs. \cite{Iyer:1986np,Konoplya:2003ii,Matyjasek:2017psv} and made even more accurate when using the Padé approximants in Refs. \cite{Matyjasek:2017psv,Hatsuda:2019eoj}.
The general WKB formula has the form \cite{Konoplya:2019hlu},
\begin{eqnarray}
\omega^2&=&V_0+A_2(\K^2)+A_4(\K^2)+A_6(\K^2)+\ldots \\\nonumber
&-& \imo \K\sqrt{-2V_2}\left(1+A_3(\K^2)+A_5(\K^2)+A_7(\K^2)+\ldots\right),
\end{eqnarray}
where $\K=n+1/2$ is a half-integer. The corrections $A_k(\K^2)$ of the order $k$ to the eikonal formula are polynomials of $\K^2$ with rational coefficients and depend on the values of higher derivatives of the potential $V(r)$ in its maximum. In order to increase the accuracy of the WKB formula, we will follow the procedure of Matyjasek and Opala \cite{Matyjasek:2017psv} and use the Padé approximants. Here, for the fundamental mode $n=0$, we will use the 13th-order WKB method with $\tilde{m}=7$, where $\tilde{m}$ is defined in Refs. \cite{Matyjasek:2017psv,Konoplya:2019hlu}, because this choice provides the best accuracy in the Schwarzschild limit and there is hope that this will be the case for more general metrics. We will see that the 13th-order WKB formula gives much better agreement with the accurate Frobenius method than the sixth and lower orders used previously in the literature for this case.

\subsection{Numerical and semianalytical calculations of quasinormal modes}

As was mentioned earlier, the mass of the black holes diminishes when $p$ is increased, achieving zero at some extremal value of $p$. Therefore, such light near extremal black holes apparently could not be created as a result of the collapse of stars. Therefore, first of all, the astrophysically motivated range of the coupling parameter $p$, that is relatively near the Schwarzschild limit, will be considered in this section.

From the accurate Frobenius data shown in Figs. \ref{fig:scal} and \ref{fig:el} for a fixed value of the radius of the event horizon $r_{+}=1$ we can see that for both scalar and electromagnetic perturbations at the lowest multipole number $\ell$ the real oscillation frequency of the fundamental mode $n=0$ is increased only by a few percent, when the coupling $p$ runs from $0.876$ until $0.92$. The damping rate is changed even less: by a fraction of one percent. Thus, the fundamental mode is almost indistinguishable from the Schwarzschild one in this range of $p$.

Though, as the mass changes with $p$, the black hole might be distinguishable from the Schwarzschild solution if one takes into account the inspiral phase.
At the same time, we can see that higher overtones are much more sensitive to variation of the coupling $p$. Thus, the $n=6$ mode changes already by a few hundred percent, which is 2  orders higher a rate of change than that of the fundamental mode. This outburst of overtones takes place mainly for the real oscillation frequency, while the damping rate is increased at an almost equidistant spacing for large $n$, so that it changes roughly proportionally to its value in the Schwarzschild limit.

From Figs. \ref{fig:4} and \ref{fig:5} we can see that near the Schwarzschild limit, the WKB method leads relatively close results at the 7th and 13th orders and is close to the sixth order results obtained in Ref. \cite{Zinhailo:2018ska}.
With the help of the Frobenius method in the limit $p \approx 0.876$ we reproduce the Schwarzschild values of quasinormal modes with great accuracy for all overtones under consideration, while for the fundamental mode
we reproduce the results of Ref. \cite{Zinhailo:2018ska} within the expected accuracy of the WKB method \cite{Schutz:1985zz,Iyer:1986np,Konoplya:2003ii,Matyjasek:2017psv,Konoplya:2019hlu} at the sixth order used there.

Summarizing the accuracy provided by different methods and in order to distinguish the inaccuracy due to the method from that coming from the parametrization of the metric we conclude that while the Frobenius method is accurate for any values of $p$ and overtone number $n$, the WKB method can only be safely used for calculation of the fundamental mode
and, and the 13th order is necessary to achieve sufficient accuracy for large values of $p$.

\section{Higher-order parametrization for the metric}

\begin{figure*}
\resizebox{\linewidth}{!}{\includegraphics{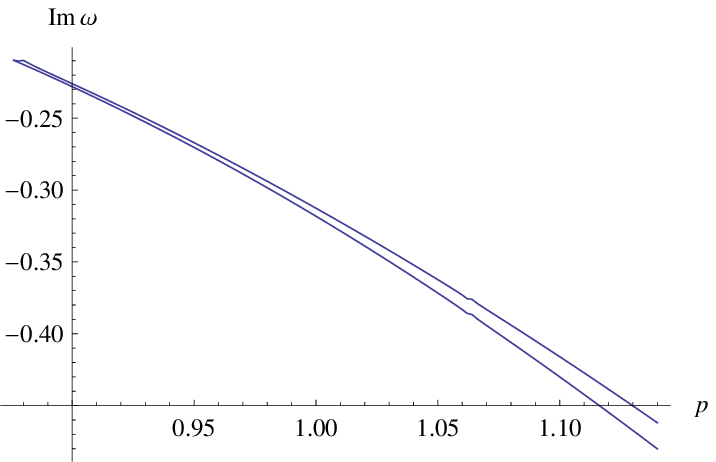}\includegraphics{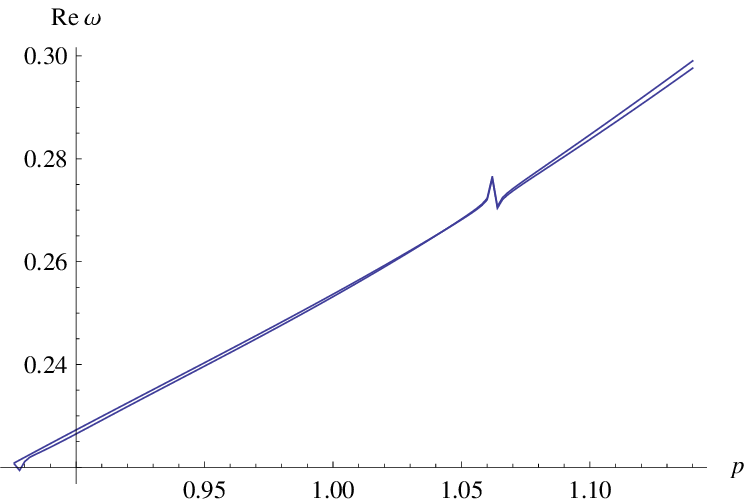}}
\caption{Real and imaginary parts of the fundamental quasinormal mode $\ell=n=0$ for scalar perturbations as a function of $p$ calculated with the help of the 13th WKB order (lower curve for $Re \omega$ and upper curve for $Im \omega$) and the Frobenius method; $\tilde{m}=7$, $r_{+} =1$. The nonmonotonic behavior near $p \approx 1.06$ is due to an insufficient number of orders (four) of the parametrization for the metric function.}
  \label{fig:4}
\end{figure*}

\begin{figure*}
\resizebox{\linewidth}{!}{\includegraphics{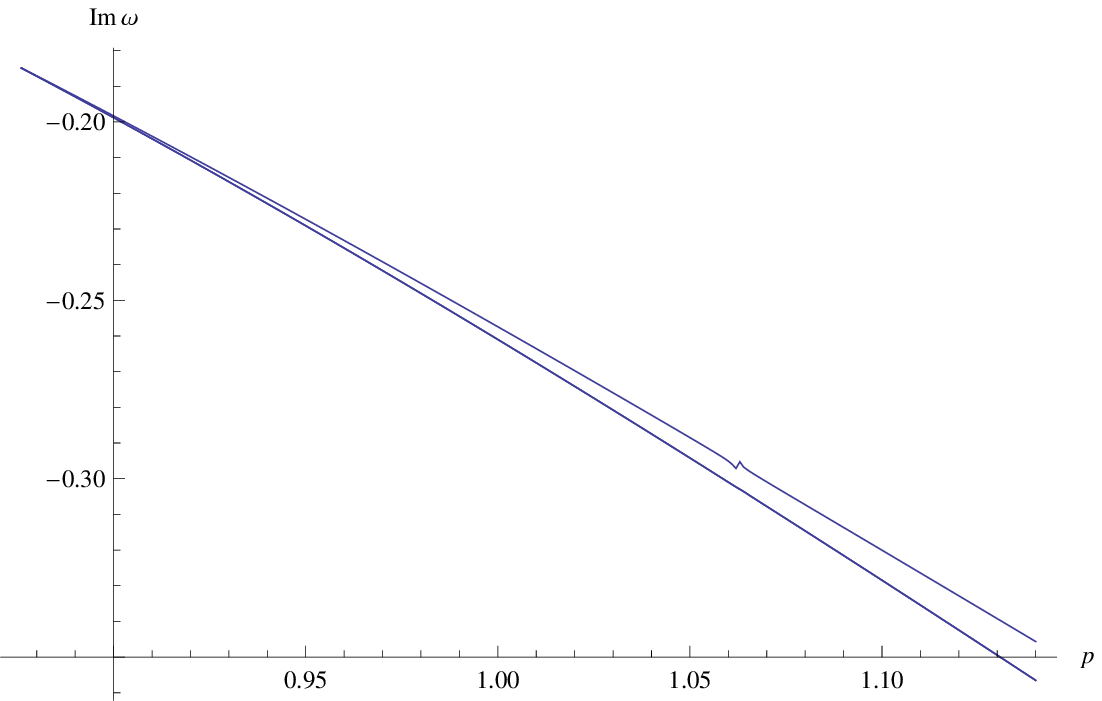}\includegraphics{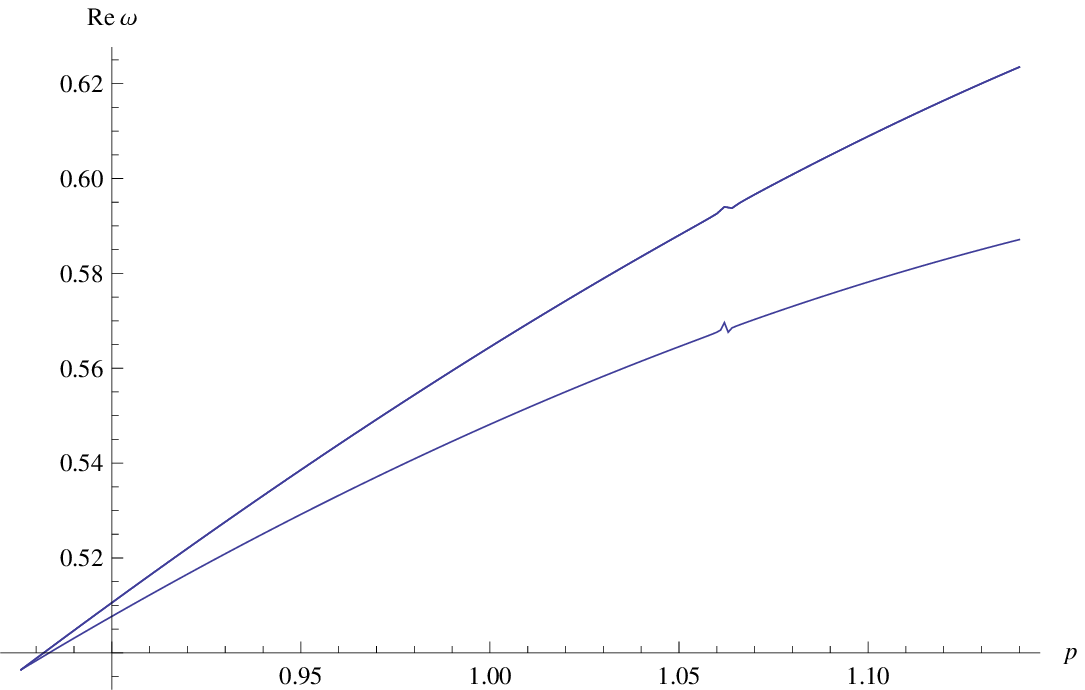}}
\caption{Real and imaginary parts of the fundamental quasinormal mode $\ell=1$, $n=0$ for electromagnetic perturbations as a function of $p$ calculated with the help of the 13th WKB order with $\tilde{m}=7$, and with the Frobenius method and 7th WKB order ($\tilde{m}=7$); $r_{+} =1$. The results of the 13th WKB order and accurate Frobenius data are very close (upper curve for $Re \omega$ and lower one for $Im \omega$) and cannot be distinguished on the plot. The nonmonotonic behavior near $p \approx 1.06$ is due to an insufficient number of orders (four) of the parametrization for the metric function.}
  \label{fig:5}
\end{figure*}

\begin{table*}
\begin{tabular}{|c|c|c|c|c|}
  \hline
  $p$ & 4th order & 5th order & 6th order  & 7th order  \\
  \hline
  \hline
  1.060 & $0.272020 - 0.383228 \imo$ & $0.275324 - 0.382319 \imo$ & $0.276697 - 0.380141 \imo$ & $0.276618 - 0.380274 \imo$ \\
  1.061 & $0.272979 - 0.384515 \imo$ & $0.275658 - 0.383442 \imo$ & $0.277033 - 0.381243 \imo$ & $0.276953 - 0.381377 \imo$ \\
  1.062 & $0.276117 - 0.385789 \imo$ & $0.275995 - 0.384565 \imo$ & $0.277371 - 0.382345 \imo$ & $0.277292 - 0.382479 \imo$ \\
  1.063 & $0.266280 - 0.381901 \imo$ & $0.276324 - 0.385698 \imo$ & $0.277701 - 0.383457 \imo$ & $0.277625 - 0.383592 \imo$ \\
  1.064 & $0.270399 - 0.386619 \imo$ & $0.276664 - 0.386822 \imo$ & $0.278042 - 0.384560 \imo$ & $0.277963 - 0.384696 \imo$ \\
  1.065 & $0.271456 - 0.388172 \imo$ & $0.277000 - 0.387952 \imo$ & $0.278042 - 0.384560 \imo$ & $0.278302 - 0.385806 \imo$ \\
  1.066 & $0.272082 - 0.389472 \imo$ & $0.277338 - 0.389083 \imo$ & $0.278718 - 0.386779 \imo$ & $0.278641 - 0.386916 \imo$\\
  \hline
\end{tabular}
\caption{The fundamental ($\ell = n =0$) mode for the scalar field calculated by the Frobenius method for the metrics which are truncation of the parametrization at different orders; $r_{0} =1$. We can see that the quasinormal modes converge with the increasing of the order and the difference between the 6th and 7th order results is a small fraction of $1\%$.}
\end{table*}

\begin{table*}
\begin{tabular}{|c|c|c|c|c|}
  \hline
  $p$ & 4th order & 5th order & 6th order  & 7th order  \\
  \hline
  \hline
  1.060 & $0.592591 - 0.300992 \imo$ & $0.592580 - 0.301161 \imo$ & $0.592699 - 0.301224 \imo$ & $0.592689 - 0.301232 \imo$ \\
  1.061 & $0.593138 - 0.301681 \imo$ & $0.593014 - 0.301842 \imo$ & $0.593134 - 0.301906  \imo$ &$0.593124 - 0.301915 \imo$ \\
  1.062 & $0.594020 - 0.302348 \imo$ & $0.593448 - 0.302524 \imo$ & $0.593568 - 0.302588 \imo$ & $0.593559 - 0.302597 \imo$ \\
  1.063 & $0.592245 - 0.302414 \imo$ & $0.593881 - 0.303208 \imo$ & $0.594001 - 0.303273 \imo$ & $0.593992 - 0.303281  \imo$ \\
  1.064 & $0.593756 - 0.303607 \imo$ & $0.594312 - 0.303889 \imo$ & $0.594434 - 0.303955 \imo$ & $0.594424 - 0.303964 \imo$ \\
  1.065 & $0.594346 - 0.304331 \imo$ & $0.594743 - 0.304573 \imo$ & $0.594865 - 0.304639 \imo$ & $0.594855 - 0.304648 \imo$ \\
  1.066 & $0.594839 - 0.305028 \imo$ & $0.5951726 - 0.305256 \imo$ & $0.595295 - 0.305323 \imo$ & $0.594993 - 0.305655 \imo$\\
  \hline
\end{tabular}
\caption{The fundamental ($\ell =1$, $n =0$) mode for the electromagnetic field calculated by the Frobenius method for the metrics which are truncation of the parametrization at different orders; $r_{0} =1$. We can see that the quasinormal modes converge with the increasing of the order and the difference between the 6th and 7th order results is a small fraction of one percent.}
\end{table*}

\begin{table*}
\begin{tabular}{|c|c|c|c|c|c|c|}
  \hline
  $p$ & $a_5$ & $b_5$ & $a_6$  & $b_6$  & $a_7$  & $b_7$ \\
  \hline
  \hline
  1.060 & $-1/3743$ & $-603/1802$ & $-301/2336 $ & $-563/1673$  & $1413/1060$ & $11/636$\\
  1.061 & $-1/5867$ & $-564/1685$ & $-533/4174$ & $-1565/4649$  & $371/274$ & $19/1099$\\
  1.062 & $-1/13431$ & $-984/2939$ & $-165/1304$ & $-461/1369$  & $337/245$ & $7/405$\\
  1.063 & $1/48062$ & $-571/1705$ & $-85/678$ & $-351/1042$  & $4/3$ & $1/57$\\
  1.064 & $1/8669$ & $-878/2621$ & $-155/1248$ & $-650/1929$  & $1535/1081$ & $12/695$\\
  1.065 & $1/4780$ & $-641/1913$ & $-475/3861$ & $-1112/3299$  & $1013/702$ & $19/1101$\\
  1.066 & $1/3308$ & $-5491/16383$ & $-343/2815$ & $-701/2079$  & $5447/3714$ & $39/2261$\\
  \hline
\end{tabular}
\caption{Higher coefficients of the metric parametrization obtained with the help of the automatic procedure developed in \cite{Kokkotas:2017zwt}; $r_{0}=1$. The lower coefficients are found as functions of $p$ in \cite{Kokkotas:2017zwt}.}
\end{table*}

Following the parametrization procedure of Ref. \cite{Rezzolla:2014mua} the new functions $A$ and $B$ are defined as follows:
\begin{eqnarray}\label{hd}
A(r)&\equiv&xh(x)\,,\\
\label{fd}
\frac{A(r)}{B(r)}&\equiv&f(x)^2\,,
\end{eqnarray}
where $x$ is the dimensionless compact coordinate
\begin{equation}
x \equiv 1-\frac{r_0}{r}\,.
\end{equation}
We write the above two functions in the following form:
\begin{eqnarray}
h(x)&=&1-\epsilon (1-x)+(a_0-\epsilon)(1-x)^2+{\tilde h}(x)(1-x)^3,\nonumber\\
\label{asympfix}
f(x)&=&1+b_0(1-x)+{\tilde f}(x)(1-x)^2,
\end{eqnarray}
where ${\tilde h}(x)$ and ${\tilde f}(x)$ are expressed in terms of the continued fractions, in order to describe the metric near the event horizon $x=0$:
\begin{align}\nonumber
{\tilde h}(x)=\frac{a_1}{\displaystyle 1+\frac{\displaystyle
    a_2x}{\displaystyle 1+\frac{\displaystyle a_3x}{\displaystyle
      1+\frac{\displaystyle a_4x}{\displaystyle
      1+\ldots}}}}\,,\\\label{contfrac}
{\tilde f}(x)=\frac{b_1}{\displaystyle 1+\frac{\displaystyle
    b_2x}{\displaystyle 1+\frac{\displaystyle b_3x}{\displaystyle
      1+\frac{\displaystyle b_4x}{\displaystyle
      1+\ldots}}}}\,.
\end{align}
At the event horizon, one has ${\tilde h}(0) = a_1, \quad {\tilde f}(0) =b_1$.

From Figs. \ref{fig:4} and \ref{fig:5} we see that at a fixed radius of the event horizon, the real oscillation frequency grows monotonically and damping rate decreases once $p$ is growing, except for a small region near $p \approx. 1.06$. There, a kind of  ``splash ''  {occurs, which is not usual for quasinormal spectra of black holes, because normally the frequencies depend on physical parameters in a smooth way.

Quasinormal modes obtained here in Figs. \ref{fig:scal} and \ref{fig:el} (for values $p$ near the Schwarzschild limit) as well as in \cite{Zinhailo:2018ska,Zinhailo:2018ska} were found for the fourth-order approximation \cite{Kokkotas:2017zwt} of the numerical solution \cite{Lu:2015cqa}. This means that $a_5 = a_6... =0$, and $b_5 = b_6... =0$. In the regime of relatively small deviation from the Schwarzschild geometry (as in figs. \ref{fig:scal} and \ref{fig:el} here) the fourth-order approximation is sufficient that is, the quasinormal modes do not change significantly if the expansion is truncated at higher orders. In other words, the convergence regime is achieved.

However, as we can see from Tables I and II, at higher $p$ in the range where the nonmonotonic dependence of $\omega$ on $p$  occurs, this is not so, and the difference between quasinormal modes calculated for the forth- and fifth-order metrics may reach a few percent. At the same time, when higher-order approximations are used according to the general automatic procedure in Ref. \cite{Kokkotas:2017zwt}, the quasinormal modes converge quickly. For example, the difference between quasinormal modes for the 5-th and 6-th order metrics is about one percent or less, while the difference between the sixth- and seventh-order is already a small fraction of $1 \%$.
We can also see that the convergence selectively  depends on $p$ in this range of parameters, being the worst for $p=1.065$, but better for smaller and larger $p$.
The coefficients of the metric parametrization $\epsilon$, $a_i$ and $b_i$ for $i \leq 4$ are found in general form, that is, as functions of $p$ and $r_{0}$ in Ref. \cite{Kokkotas:2017zwt}, while higher-order coefficients can be seen in Table III here for particular values of $p$ considered in Tables I and II. Using the above data in \cite{Kokkotas:2017zwt} and Table~III one can obtain quasinormal modes for other spin fields and multipole numbers $\ell$.

\section{Conclusions}

The present paper is revising the earlier studies of quasinormal modes of non-Schwarzschild asymptotically flat black holes in the Einstein-Weyl gravity in two aspects: accurate calculation of overtones which cannot be found with the WKB or time-domain integration methods and testing the convergence of the parametrization for the metric.  We have found that:
\begin{itemize}
\item Here we have accurately calculated quasinormal modes of the Einstein-Weyl black hole by two alternative methods which achieve the regime of convergence and are very well in concordance with each other. This way we have shown that the discrepancy between Refs. \cite{Zinhailo:2019rwd,Zinhailo:2018ska} and Ref. \cite{Cai:2015fia} is resolved in the favor of Ref.  \cite{Zinhailo:2019rwd,Zinhailo:2018ska} at small coupling, while at larger coupling the results of \cite{Zinhailo:2019rwd,Zinhailo:2018ska} are corrected. In other words, the accurate quasinormal modes obtained here are in a very good agreement with Refs. \cite{Zinhailo:2019rwd,Zinhailo:2018ska} in the regime of small coupling and significantly differ from \cite{Cai:2015fia}.
\item The first several overtones are very sensitive to the change of the coupling constant. While the fundamental mode is changed only within a couple of percent, the higher overtones can change by hundreds of percent. This may be important for reproduction of the quasinormal ringing at earlier times \cite{Giesler:2019uxc}, for the so-called mode instability \cite{Jaramillo:2020tuu} and for testing the geometry of the event horizon \cite{Konoplya:2022pbc}. In Ref. \cite{Jaramillo:2021tmt} it is shown that the overtone outburst may already occur within the near-future detection range while in Ref. \cite{Oshita:2022yry} the possibility of the detection of overtones with LISA is argued.
\item The unusual nonmonotonic dependence of the quasinormal frequencies upon the coupling parameter $p$ which was found in earlier publications \cite{Zinhailo:2019rwd,Zinhailo:2018ska} is the result of the insufficient number of terms at which the metric parametrization was truncated, when approximating the numerical solution by an analytical expression. Extension to higher orders allows one to achieve convergence when computing quasinormal modes.
\end{itemize}

 Our work could be extended in a number of ways, first of all, by consideration of the Dirac and gravitational perturbations. The latter would be a highly challenging problem, because the linearization of perturbation equations in the higher-derivative gravity will  most probably lead to a system of chained differential equations of higher than the second order. Nevertheless, the phenomenon of sensitivity of the overtones found here is connected with the near-horizon deformation of the geometry of the non-Schwarzschild black hole \cite{Konoplya:2022pbc,Konoplya:2022hll}. Therefore, qualitatively the same behavior of overtones must take place for perturbations of other spin fields as well.

\acknowledgments
The author would like to thank the support of the Czech Science Foundation (GAČR) under Grant No. 19-03950S and A. Zhidenko for most useful discussions.

\end{document}